\documentclass[sigconf]{acmart}
\usepackage{hyperref}
\usepackage{hyperxmp}
\usepackage{subfigure}
\usepackage{algorithm}
\usepackage{algorithmic}

\begin{document}
\settopmatter{printacmref=false} 
\renewcommand\footnotetextcopyrightpermission[1]{} 
\pagestyle{plain} 

\title{WiCross: Indoor Human Zone-Crossing Detection Using Commodity WiFi Devices}

\author{Weiyan Shi}
\affiliation{%
  \institution{Peking University}
  \city{Beijing}
  \country{China}}
\email{shiweiyan@stu.pku.edu.cn}

\author{Xuanzhi Wang}
\affiliation{%
  \institution{Peking University}
  \city{Beijing}
  \country{China}}
\email{xuanzhiwang@stu.pku.edu.cn}

\author{Kai Niu}
\affiliation{%
  \institution{Peking University}
  \institution{Beijing Xiaomi Mobile Software Company Ltd.}
  \city{Beijing}
  \country{China}}
\email{xjtunk@pku.edu.cn}

\author{Leye Wang}
\affiliation{%
  \institution{Peking University}
  \city{Beijing}
  \country{China}}
\email{leyewang@pku.edu.cn}

\author{Daqing Zhang}
\affiliation{%
  \institution{Peking University, Beijing}
  \institution{Telecom SudParis, Evry}
  \country{China/France}}
\email{dqzhang@sei.pku.edu.cn}
\authornote{Corresponding author. This is a preprint of the paper accepted at Ubicomp/ISWC 2023 Poster. The final version will be available in the ACM Digital Library.}

\renewcommand{\shortauthors}{Shi, et al.}

\begin{abstract}
Detecting whether a target crosses the given zone (e.g., a door) can enable various practical applications in smart homes, including intelligent security and people counting. The traditional infrared-based approach only covers a line and can be easily cracked. In contrast, reusing the ubiquitous WiFi devices deployed in homes has the potential to cover a larger area of interest as WiFi signals are scattered throughout the entire space. By detecting the walking direction (i.e., approaching and moving away) with WiFi signal strength change, existing work can identify the behavior of crossing between WiFi transceiver pair. However, this method mistakenly classifies the turn-back behavior as crossing behavior, resulting in a high false alarm rate. In this paper, we propose WiCross, which can accurately distinguish the turn-back behavior with the phase statistics pattern of WiFi signals and thus robustly identify whether the target crosses the area between the WiFi transceiver pair. We implement WiCross with commercial WiFi devices and extensive experiments demonstrate that WiCross can achieve an accuracy higher than 95\% with a false alarm rate of less than 5\%.
\end{abstract}



\maketitle

\section{Introduction}

With the popularity of various smart spaces, both academia and industry are paying increasing attention to smart services such as intelligent security~\cite{kodali2016iot, somani2018iot} and people counting. To ensure the security of smart indoor environments, detecting whether people pass through the door plays a significant role. The existing work utilizes cameras to detect whether people pass through the door. But cameras face serious privacy concerns in indoor environments, especially in bathrooms and bedrooms~\cite{grau2015can}. On the other hand, due to the good privacy protection characteristics, WiFi-based contactless sensing has been widely explored for intrusion detection~\cite{li2017ar,li2020wiborder}, activity recognition~\cite{gao2021towards,wang2016rt}, and indoor localization~\cite{abbas2019wideep,wu2021witraj}.  Thus, it is attractive to employ the ubiquitous commercial off-the-shelf (COTS) WiFi devices to detect people passing through doors.

To determine whether people pass through doors, existing detection methods can be grouped into two categories: (1) model-based methods and (2) pattern-based methods. 
The model-based method~\cite{li2020wiborder} places the transceivers inside the wall such that the strength of WiFi channel state information (CSI) induced by a person inside the wall is much higher than that induced by a person outside the wall. Then a threshold is set to identify whether the person has crossed the door with the strength of received CSI signals. However, the placement of the devices is complicated and not applicable to scenes without walls. The pattern-based method~\cite{yang2019door} takes the phase difference of CSI signals as input to train a convolutional neural network (CNN) and extracts the walking direction of the person, which is further employed to determine whether the person crosses the door. However, it fails to identify the case where a person approaches the door and then returns.

In this paper, we observe that when a person crosses the LoS path between the WiFi transmitter  and receiver, the phase information of the received CSI signal exhibits a distinct pattern. More importantly, we find that the phase information of received CSI signals presents distinguishable patterns when the target turns back from the LoS. Thus, compared to the pattern-based method~\cite{yang2019door}, we can differentiate the behaviors of crossing and turning back. 
The above observation indicates that by only leveraging a pair of a transmitter and a receiver, we can distinguish whether a person has passed through the door according to whether the person has passed through the LoS. Based on this idea, we implement WiCross, a system for identifying whether a person crosses the LoS using commercial WiFi devices. WiCross is able to detect the behavior of a person passing directly through the door and can distinguish it from a person approaching the door and turning back. Extensive experiments under various conditions (i.e., different locations, orientations, LoS distances) illustrate WiCross can achieve an accuracy higher than 95\% with a false alarm rate of less than 5\%.

\section{WICROSS SYSTEM}

\subsection{System Setup}
\begin{figure}[h]
  \centering
  \includegraphics[width=0.8\linewidth]{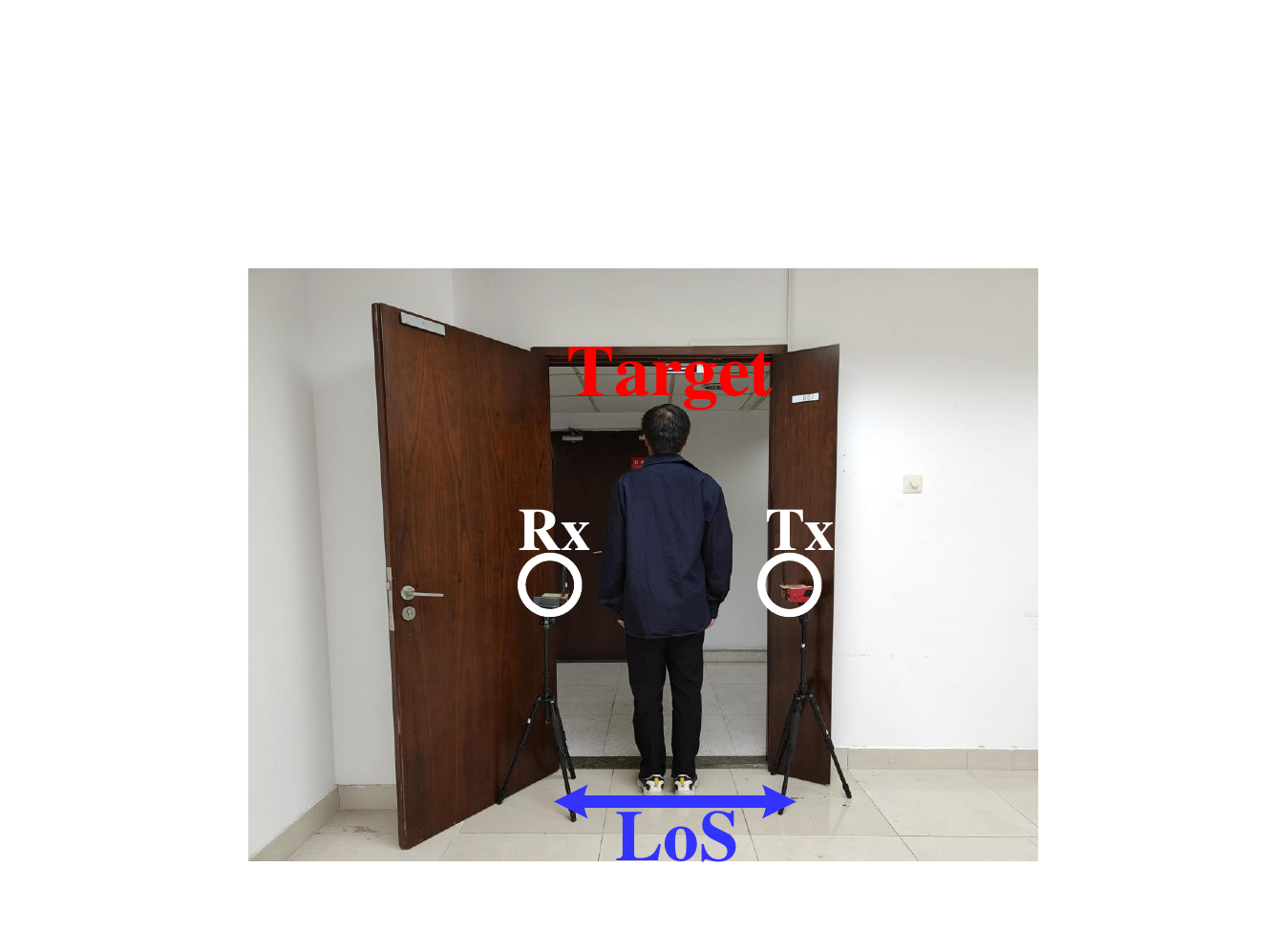}
  \caption{Experimental setup}
  \label{Figure 1}
\end{figure}

WiCross prototype employs two Gigabyte Mini-PCs equipped with the cheap off-the-shelf Intel 5300 WiFi card as transceivers. The transmitter (Tx) is installed with one omnidirectional antenna, while the receiver (Rx) is configured with three omnidirectional antennas. WiCross operates on the 5.24GHz frequency with a bandwidth of 40MHz. The receiver continuously collects CSI packets from the transmitter at a sampling rate of 1000Hz. 
Figure~\ref{Figure 1} illustrates the transceiver's placement scene. The transceivers are deployed on opposite sides of the doorway. The LoS distance between them depends on the length of the doorway. And the height is set as 1.1m to suit most persons as soon as possible.

\subsection{System Overview}

The WiCross system consists of four key modules:

\begin{enumerate}
    \item \textbf{Data acquisition.} In WiCross, CSI readings are collected from the WiFi receiver in real-time and then sent to a laptop using a local network. The laptop configured with an Intel Core i7 7700U CPU and 16GB memory runs the algorithm implemented by Matlab to process the data.
 
    \item \textbf{Data processing.} WiCross first utilizes a moving average filter with a sliding window of 50 to denoise the raw CSI signals.
    To further reduce the phase noises induced by central frequency offset (CFO) and sampling frequency offset (SFO) due to the unsynchronized clocks between transceivers, WiCross leverages the CSI ratio between the different antennas in the same receiver~\cite{zeng2019farsense,primer2021boosting}. Finally, WiCross automatically segments the signal when the target crosses the doorway using the Automatic Gain Control (AGC)~\cite{jimenez2004design} value by a threshold-based method.
    
    \item \textbf{Pattern extracting.} After obtaining the signal segment, WiCross calculates the phase information from the CSI ratio signals and generates the accumulated phase patterns.
    When the person approaches the LoS path, the cumulative phase increases. If the person is just on the LoS path, the cumulative phase reaches its maximum value. When the person moves away from the LoS path, the cumulative phase decreases. The following section will explain the basic idea of judging whether a person crosses the LoS path using the cumulative phase, as well as how to obtain the cumulative phase patterns.

    \item \textbf{Behavior detection.} WiCross determines whether a person crosses a door by examining the cumulative phase pattern's local maxima and minima. If there is one local maximum and two local minimums on either side, the person crosses the door. Otherwise, the person is not considered to cross the door.
\end{enumerate}

\section{KEY TECHNIQUE}

\begin{figure}[h]
  \centering
  \includegraphics[width=0.7\linewidth]{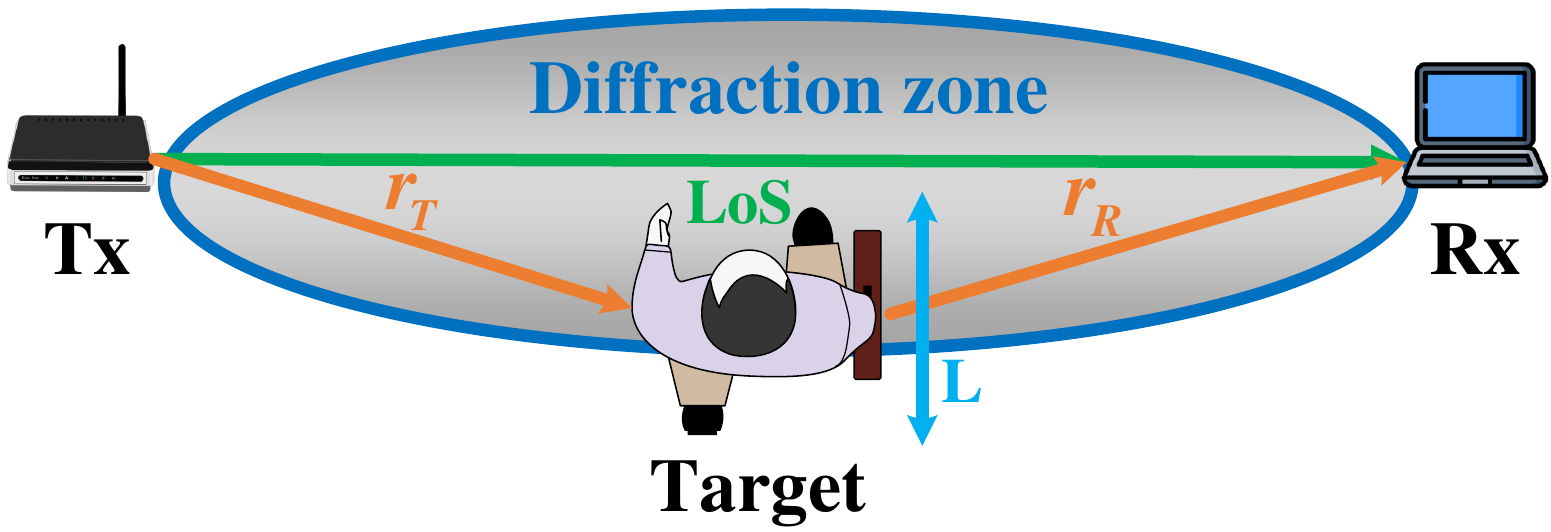}
  \caption{Illustration of  wireless signals' propagation process.}
  \label{Figure 2}
\end{figure}

Based on the diffraction sensing theory \cite{xz2023wimeasure}, when a finite target locates at the diffraction zone (gray area in Figure~\ref{Figure 2}), the CSI signal can be expressed as

\begin{equation}
\begin{split}
H&=H_{LoS}+H_{Target}\\
&=H_{LoS}+\left \{ -\frac{i}{2\lambda }\frac{E_{0}e^{i\varphi _{0}}}{\sqrt{4\pi }}\int_{L}^{}\frac{e^{-\frac{i2\pi (r_{T}+r_{R}))}{\lambda }}}{r_{T}r_{R}} dL^{'} \right \} ,
\end{split}
\label{eq1}
\end{equation}
where $L$ represents the length of the reflector $target$ and $\lambda$ represents the wavelength of the wireless signals. $E_{0}$ and $\varphi_{0}$ denote transmit energy and initial phase, respectively. The distances from the target to the transmitter and receiver are represented by $r_{T}$ and $r_{R}$, respectively. Assuming that the reflector $target$ is moving, then the received wireless signals consist of both static propagation paths ($H_{LoS}$) and dynamic propagation paths ($H_{Target}$).

\begin{figure}[h]
    \centering
    \subfigure[]{
    \includegraphics[width=0.45\linewidth]{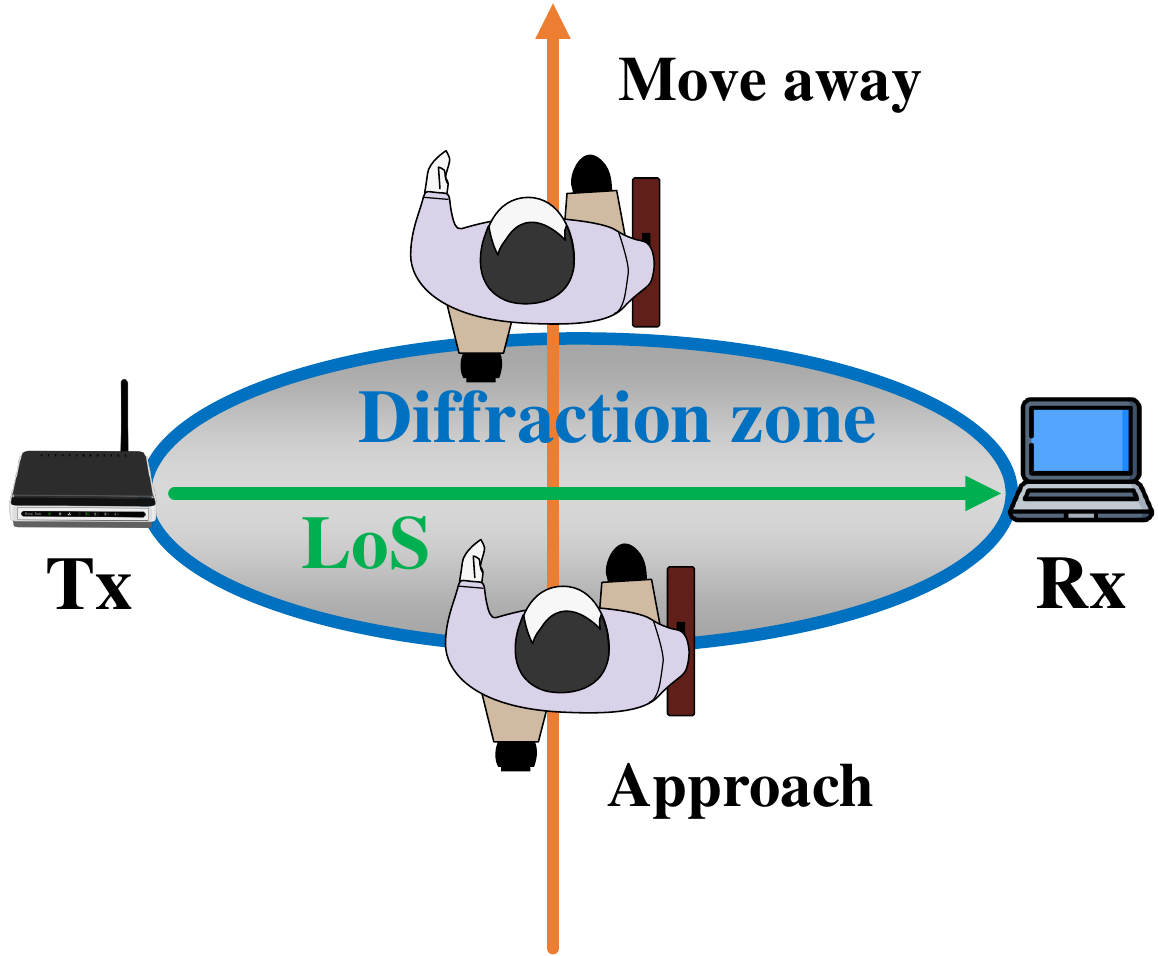}}
    \subfigure[]{
    \includegraphics[width=0.45\linewidth]{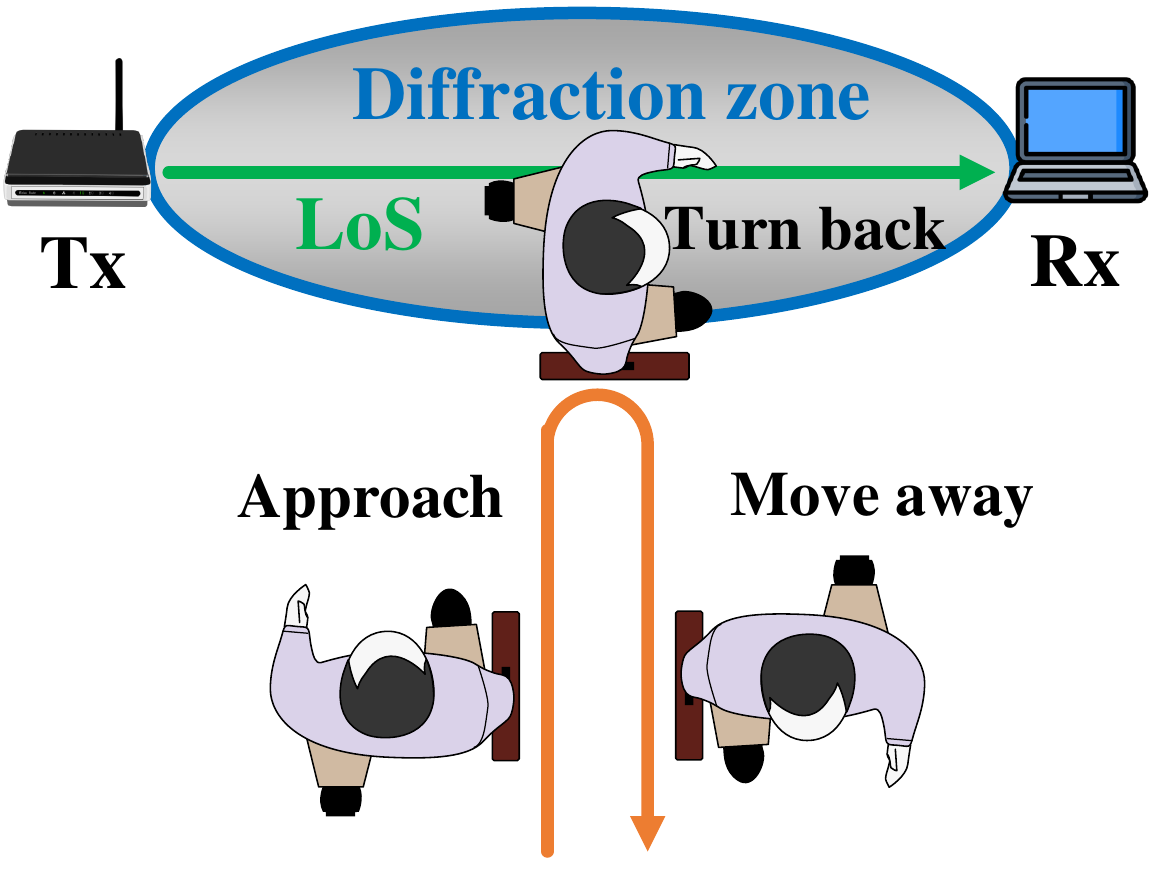}}
    \subfigure[]{
    \includegraphics[width=0.45\linewidth]{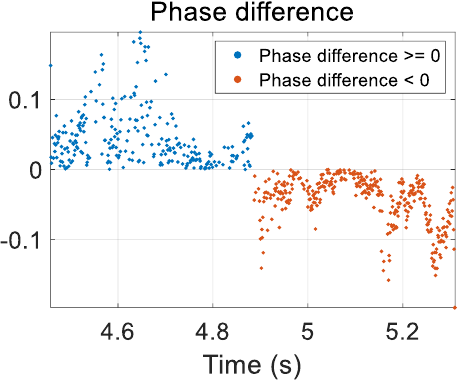}}
    \subfigure[]{
    \includegraphics[width=0.45\linewidth]{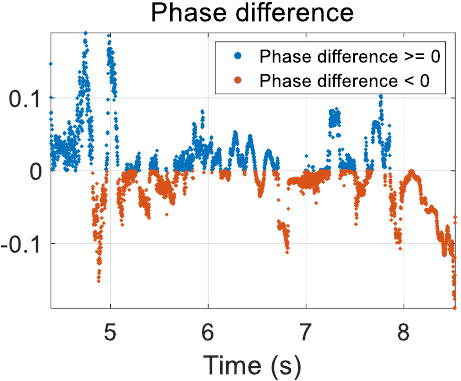}}
    \subfigure[]{
    \includegraphics[width=0.45\linewidth]{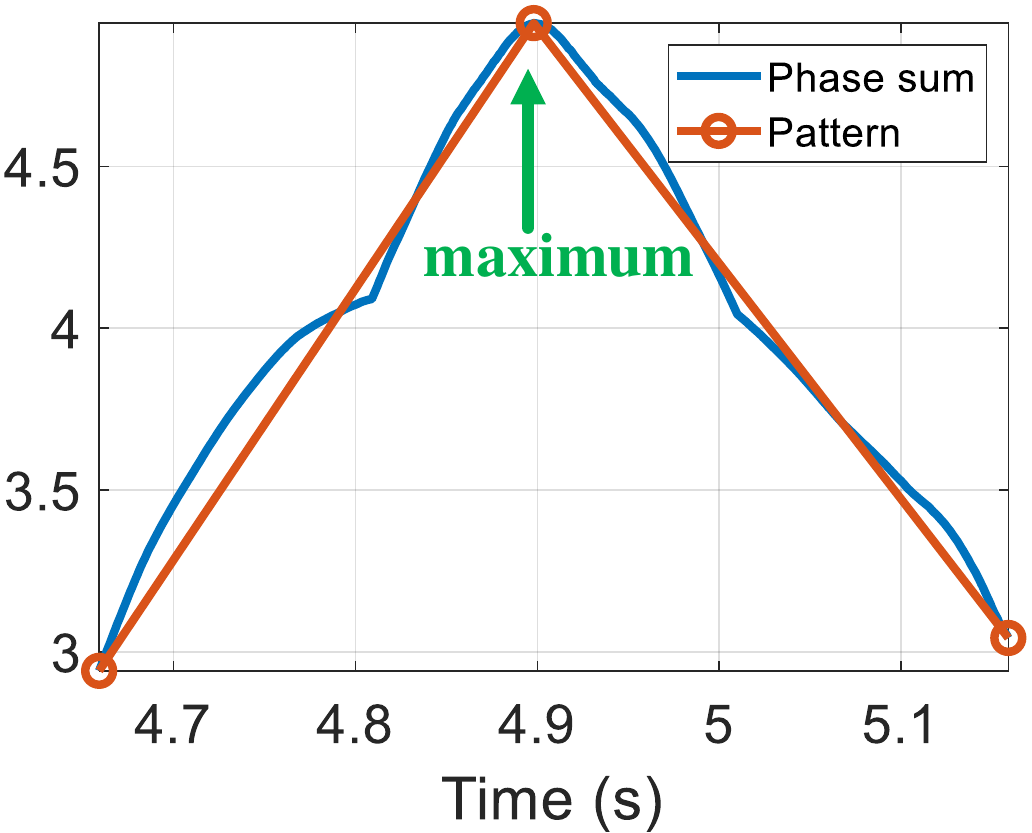}}
    \subfigure[]{
    \includegraphics[width=0.45\linewidth]{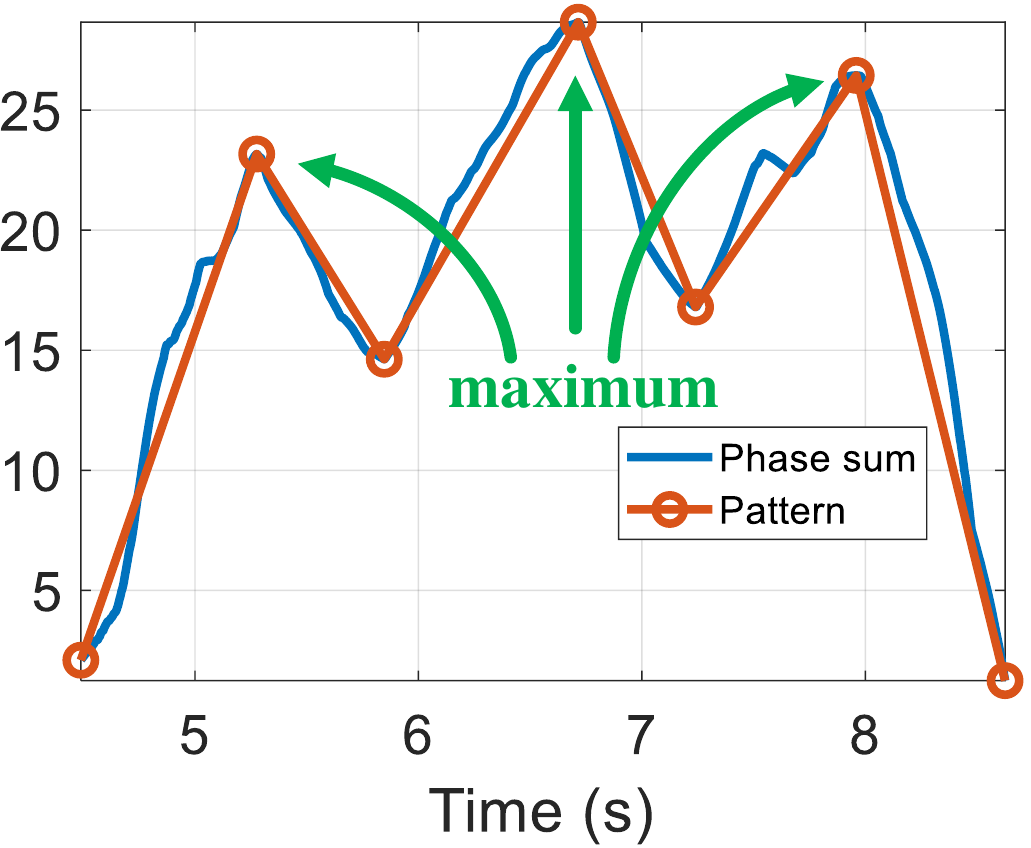}}
    \caption{Key observations.}
    \label{Figure 3}
\end{figure}

According to Equation~\ref{eq1}, the phase of the CSI signal is a linear function of the sum of the distances between the person and the transceiver. Thus, we can conclude that the phase of the CSI signal increases when the subject is close to the LoS and decreases when the target moves far from the LoS. For instance, Figure~\ref{Figure 3}a illustrate the target crosses the diffraction zone, while Figure~\ref{Figure 3}c and Figure~\ref{Figure 3}e give the corresponding phase difference and pattern. Hence, we can determine if a person crosses the LoS path by analyzing the CSI signal phase pattern. In contrast, when the target turns back into the diffraction zone (Figure~\ref{Figure 3}b), the phase will experience more process of changes (Figure~\ref{Figure 3}d and Figure~\ref{Figure 3}f). With the distinguishable phase change patterns, we can robustly identify the behavior of crossing diffraction zone with low a false alarm rate.

Algorithm for pattern extracting (Alg~\ref{alg1}) starts by computing the initial phase of the CSI signal by taking the angle of the difference between consecutive elements. Subsequently, the phase difference between adjacent elements is calculated. A cumulative sum of these phase differences is then computed, allowing the algorithm to identify potential crossing points.

To determine the presence of a crossing event, the algorithm examines the number of local maxima in the phase sum array. If a single local maximum is detected, it indicates a crossing event. Otherwise, either multiple local maxima or the absence of local maxima suggests a non-crossing scenario.

\begin{algorithm}
	\renewcommand{\algorithmicrequire}{\textbf{Input:}}
	\renewcommand{\algorithmicensure}{\textbf{Output:}}
	\caption{Algorithm for Pattern Extracting}
	\label{alg1}
        \begin{algorithmic}[1]
            \REQUIRE Processed CSI signal: $CSI$
            \ENSURE Crossing
            
            \state $phase \leftarrow \angle(\Delta CSI)$
            
            \state $\Delta phase \leftarrow phase[i+1] - phase[i], \forall i \in [1, \text{length}(phase)-1]$
            
            \state \textbf{for} $i\leftarrow 2  \ \textbf{to} \ \text{length}(\Delta phase) $ 
            
            \state  $  \quad phase\_sum[i] \leftarrow phase\_sum[i-1] + \Delta phase[i]$
            
            \state \textbf{end for}
            
            \state $N = \{\text{length}(s_i) \mid s_i \text{ is a local maximum in }phase\_sum\}$
            
            \state \textbf{if} {$N = 1$} \ \textbf{then}
            
            \state $\quad \text{Crossing} \leftarrow True $
            
            \state \textbf{else}
            
            \state $\quad\text{Crossing} \leftarrow False$
            
            \state \textbf{end if}
            
        \end{algorithmic}
\end{algorithm}

The pattern is unique and stable when the person crosses the door, regardless of their location and directions. Moreover, the consistency of the pattern is maintained for different users without the need to calibrate the system for each individual. WiCross could achieve good performance without training by employing such a pattern.

\section{EVALUATION}
\begin{figure}[h]
  \centering
  \includegraphics[width=0.5\linewidth]{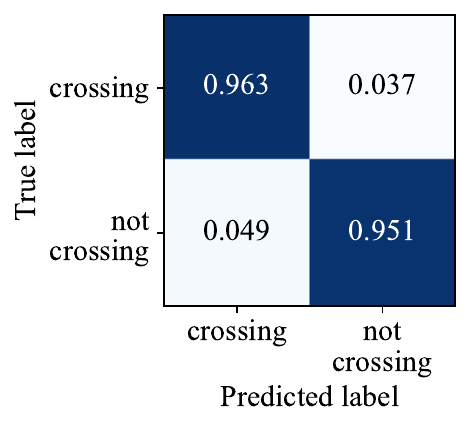}
  \caption{Confusion matrix}
  \label{Figure 4}
\end{figure}

We conduct 816 experiments, with 409 groups for people who cross the door and 407 groups for those who do not. In the experiments where the door is not crossed, there are 209 groups where people walk close to the door and return back, and there are 198 groups where people walk near the door. As shown in Figure~\ref{Figure 4}, WiCross achieves a 0.957 accuracy for people passing through doors and a false alarm rate of 4.9\%. 

We then evaluate WiCross' robustness in the context of different lines of sight, different positions, and different directions through the door.

\subsection{Impact of LoS}
Figure~\ref{Figure 5}a illustrates the recognition accuracy of individuals crossing LoS with four different LoS distances. As LoS distance increases, accuracy decreases because sensing ability weakens due to signal energy reduction~\cite{wang2022placement}. In spite of this, WiCross is capable of achieving high levels of accuracy under four different lengths of LoS path.

\subsection{Impact of User Diversity}
\begin{figure}[h]
    \centering
    \subfigure[]{
    \includegraphics[width=0.45\linewidth]{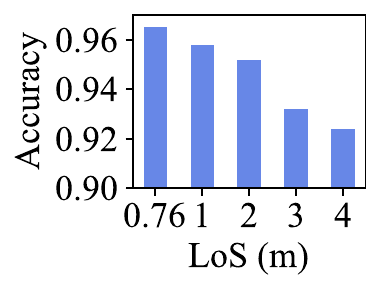}}
    \subfigure[]{
    \includegraphics[width=0.45\linewidth]{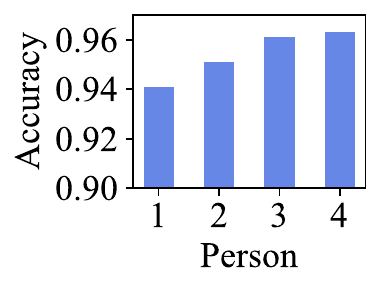}}
    \caption{Impact of LoS and user diversity on WiCross accuracy.}
    \label{Figure 5}
\end{figure}

A total of four volunteers are recruited to cross the door. Figure~\ref{Figure 5}b shows there are no noticeable differences in pattern recognition accuracy between volunteers, ranging between 0.94 and 0.96. As Volunteer 1 has the largest body size, the accuracy rate is the lowest. Due to the increasing width of the person, the signal becomes more complex~\cite{xz2023wimeasure}, resulting in a decrease in accuracy as the width increases.
\subsection{Impact of Direction and Position of the Person Crossing}
\begin{figure}[h]
    \centering
    \subfigure[]{
    \includegraphics[width=0.45\linewidth]{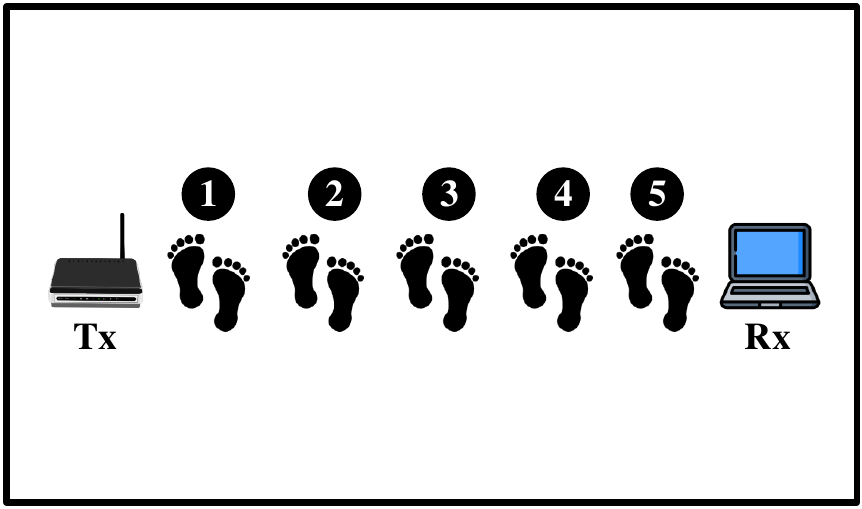}}
    \subfigure[]{
    \includegraphics[width=0.45\linewidth]{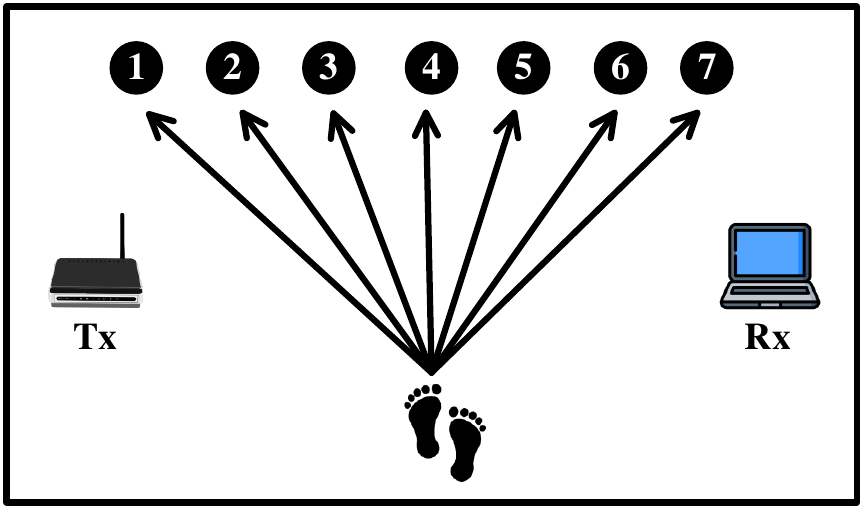}}
    \subfigure[]{
    \includegraphics[width=0.45\linewidth]{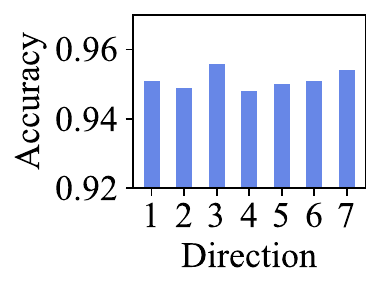}}
    \subfigure[]{
    \includegraphics[width=0.45\linewidth]{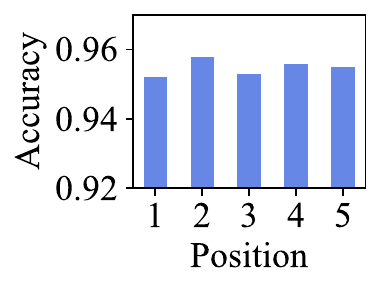}}
    \caption{Impact of direction and position diversity on WiCross accuracy.}
    \label{Figure 6}
\end{figure}

We conduct experiments of the person crossing the door at five different positions (Figure~\ref{Figure 6}a) and in seven different directions (Figure~\ref{Figure 6}b). WiCross is capable of achieving high recognition accuracy in all five directions (Figure~\ref{Figure 6}c) and seven directions (Figure~\ref{Figure 6}d). Therefore, the direction and position of the crossing do not affect the accuracy of the crossing.

\section{CONCLUSION}
Using the wireless sensing diffraction model, we successfully implement a system to robustly detect human crossing-door behavior. We can accurately identify whether a person is crossing a door without the need for model training. In the home environment, the system displays a high degree of accuracy when people pass through the door. The proposed system has broad application prospects in applications such as indoor counting and can enable intelligent security.

\begin{acks}
This research is supported by NSFC A3 Project 62061146001, PKU-NTU Collaboration Project, and the Project funded by China Postdoctoral Science Foundation (No. 2021TQ0048).
\end{acks}

\bibliographystyle{ACM-Reference-Format}
\bibliography{main}

\end{document}